\documentclass[aps,prl,superscriptaddress,amsmath,amssymb,showpacs]{revtex4}
\usepackage{amsmath}
\usepackage{graphicx}
\usepackage{amssymb}
\usepackage{subfigure}
\usepackage{epstopdf}
\usepackage{dcolumn}
\usepackage{bm}

\newcommand{\nd}{N_{DNA}}
\newcommand{\nw}{  N_{H_2O}}
\newcommand{\nt}{N_{Tot}}
\newcommand{\hb}{\frac{3\hbar^2}{2M}}

\makeatletter

\begin{document}

\title{Changes in the zero point energy of the protons as the source of the binding energy of water to A phase DNA  }

\author{G. F. Reiter}
\affiliation{Physics Department,
  University of Houston,
  Houston, TX 77204}
  \author{R. Senesi}
\affiliation{Dipartimento di Fisica and Centro NAST, Universit\`a degli Studi di Roma "Tor Vergata", Via della Ricerca Scientifica 1, 00133
Roma, Italy}

\author{J. Mayers}
\affiliation{ISIS, Rutherford Appelton Laboratory, Chilton, Didcot, England}

\date{\today}
\begin{abstract}
\noindent The zero point kinetic  energy of protons in water is large on the scale of chemical interaction energies( ~29 Kj/mol in bulk room temperature water). Its value depends upon the structure of the hydrogen bond network, and can change as the network is confined or as water interacts with surfaces. These changes have been observed to be large on a chemical scale for water confined in carbon nanotubes and in the pores of xerogel, and may play a fundamental, and  neglected,  role in biological processes involving confined water. We measure the average momentum distribution of the protons in salmon Na-DNA using Deep Inelastic Neutron Scattering, for a weakly hydrated (~6w/bp) and a dehydrated fiber sample. This permits the determination of the change in total kinetic energy of the system per water molecule removed from the DNA and placed in the bulk liquid. This energy is  equal, within errors,  to the measured enthalpy for the same process, demonstrating that changes in the zero point motion of the protons, arising from changes in structure as water molecules are incorporated in the DNA, are a significant factor in the energetics of the transition from the A to B phase with hydration, in this case, providing the entire binding energy of the water molecules to the DNA. The shape of the momentum distribution in the dehydrated phase is consistent with coherent delocalization of  some of the protons in  a double well potential, with a separation of the wells of ~.2$\AA$.  

\end{abstract}
\maketitle
Chemical interactions occurring in bulk water typically represent small changes in the energy of constituents  compared to the energy sequestered in the zero point motion of the protons in the water, primarily in that of the stretch mode.  To the extent that this energy does not change, it may be ignored, and usually is. Nearly all simulations of water in biological systems are done with models of water for which these changes cannot occur. The energy does change, however,  as the structure of the hydrogen bond network changes.   With the advent of Deep Inelastic Neutron Scattering  measurements\cite{Andreani05}, it has become possible to observe the changes in zero point kinetic energy directly.  These changes can be large($\geq$KT)and in either direction. Water confined in carbon nanotubes has a transition in which the kinetic energy of the protons in each water molecule is reduced by ~6Kj/mol\cite{nanoprl} , while water confined in the smallest pores of xerogel the kinetic energy increases by  ~14Kj/mol\cite{Garb1}.      We present here a system where these changes play a significant, indeed, the dominant role,  in an important and well studied biological process, the transformation of DNA from the A to the B phase with the absorption of water.

  DNA is hydrophylic, changing its structure as it incorporates water molecules,  transforming continuously from the A phase to the biologically active B phase as the number of hydrated water molecules increases from 1 to about 20.\cite{Albiser:2001p1002} The energy input necessary to remove a molecule of water at any level of hydration and place it in bulk water can be measured with Differential Scanning Calorimetry. It varies with the level of hydration and the composition of the DNA, and is on the order of 10kJ/mol/water molecule, decreasing with increasing numbers of hydrating waters.\cite{virnik:2001p1026} The attraction that leads to the hydrophylicity, and the source of the energy that produces the deformation of the A structure to become the B structure,  is usually thought  of as primarily electrostatic. \cite{Mazur:2008p1237} However, as , as the configuration of the surroundings of the water molecule change, the potential energy surface of the proton changes. As a consequence, the  ground state wave function changes, and  with that there is a concomitant change in the zero point kinetic energy of the proton. This can be observed indirectly in the red shift of the vibrational modes from the gas phase to the bulk liquid,  where this   change in zero point energy  accounts for ~25$\%$ of the enthalpy difference between the two phases at room temperature.   It has recently become possible to measure accurately the kinetic energy of protons in water as well as other  hydrogen bonded condensed matter systems.\cite{Andreani05} This energy is almost entirely due to the zero point motion of the protons.  The temperature plays a relatively small role, as the kinetic energy depends primarily  on the degree of localization of the protons in the structure surrounding them. As the water of hydration is incorporated into DNA, the structure surrounding  both the protons in the water and those in the DNA changes,  and we would expect changes in the kinetic energy of the protons involved.  We show here that reductions in the kinetic energy of the zero point motion of the protons in weakly hydrated salmon DNA, containing ~6 water molecules/base pair, are not only significant, they  account for the entirety of the enthalpy of  hydration of the molecules, at that level of hydration. Changes in the zero point energy of the protons in the DNA-water  complex are therefore a major factor in the transformation of the A to the B phase of DNA, at least at these relatively low levels of hydration.  There is also evidence, in the shape of the momentum distribution,  that some of the hydrogen bonds in the dry DNA, have a double well character, as revealed by a characteristic oscillation in the momentum distribution.  This property is lost upon hydration.

The kinetic energy of the protons in the DNA-water complex is readily measured using Deep Inelastic Neutron Scattering(DINS). This method has been described in detail elsewhere. \cite{Andreani05} We briefly summarize it here. At high energy transfers, the usual  neutron scattering function S($\vec{q},\omega$), which gives the differential crossection for scattering, is described by the impulse approximation limit, in which the target particle behaves as a free particle for the duration of the scattering process. It is given by

\begin{equation}
S(\vec{q},\omega)=\int n(\vec{p})\delta(\omega-\frac{\hbar q^2}{2M}-\vec{q}\cdot\frac{\vec{p}}{ M})d\vec{p}
\end{equation}
where n($\vec{p}$) is the momentum distribution of the protons, $\hbar\omega$ and $\hbar\vec{q}$ the transferred energy and momentum, respectively, and M the mass of the proton. S(q,$\omega$) is thus the Radon transform of 
n($\vec{p}$), and as the transform is invertible, n($\vec{p}$) can be obtained directly from the scattering  data. This inversion is accomplished by a method \cite{RS} in which the momentum distribution is represented, for samples such as ours that are  isotropic , as 
a series expansion
\begin{equation}
\label{defnp}
n(p)=\frac{e^{-\frac{p^{2}}{2\sigma^{2}}}}{(\sqrt{2\pi}\sigma)^3}
\cdot\Big[1+\sum_{n=2}^{\infty}a_n (-1)^nL^{\frac{1}{2}}_{n}(\frac{p^2}{2\sigma^2}) \Big]
\end{equation}
where $L^{\frac{1}{2}}_{n}(\frac{p^2}{2\sigma^2})$ are the associated Laguerre polynomials, and 
the parameter $\sigma$  is related to the kinetic energy of the proton by K.E.=$\frac{3\hbar^2}{M}\sigma^2$.\cite{Brazil} The coefficients $a_n$ are determined by a least squares fit of the data for  $S(\vec{q},\omega)$ The instrument used was Vesuvio, at ISIS, the pulsed neutron source at the Rutherford Laboratory in England. The instrument has 64 detectors, arranged with scattering angles from 35 to 75 deg on both sides of the beam. Since in the impulse approximation limit, the data is determined entirely by the scaled variable $y=\frac{M}{q}\left(  \omega - { \frac{\hbar q^2}{2M}} \right)$, all detectors give the same information when the time of flight data is converted to the y variable. The parameters of the fit are obtained by simultaneously fitting the data from all the detectors. The consistency of the data can be checked by comparison of the rescaled data from individual detectors, which all collapse onto a single curve. 

The measurements were made on fibrous salmon Na-DNA  obtained from Sigma. The fibers were mechanically broken up into segments ~2mm in size to insure a random orientation in the sample.  The total sample, approximately 5 grams,  was weighed and  packed into an aluminum container. No special attempt had been made to control the humidity of the sample environment,  which had been  the local atmosphere for some weeks.  The container was sealed with an indium seal, and placed in  the neutron beam in an evacuated chamber.    After obtaining the desired number of counts, the sample was removed from the container, weighed in air, and placed in a vacuum at 40 degrees Centigrade for 24 hours.  The sample was reweighed, again in air, and packed into the same container and resealed.  The weighing  process took approximately 20min. After counting for a day, the sample was removed and promptly reweighed. The difference between the weight upon removal from the dehydrating oven and upon removal from the sample holder was .012gr, presumably due to  hydration during the period the sample was being packed into the sample holder or being weighed the second time, indicating that the protocol was sufficiently accurate as to the weight of the sample for our purposes. The weight of the initial hydrated sample was 4.76gr, that of the dehydrated sample as removed from the sample holder 4.16gr. We will use these values to calculate the number of water molecules/nucleotide removed by the dehydration process. This is accomplished using the known stoichiometry of the Na-Dna salt. We used the value for the ratio of G-C to A-T of .436.\cite{marmur, hua} Our results are very insensitive to this ratio , as the weight ratio of protons to heavier atoms in the G-C and A-T pairs is nearly the same. The number of water molecules  removed is ~6/nucleotide, which is sufficient to produce a substantial change in the structure of the DNA. \cite{Lee:2004p1013} This is the number expected for the typical levels of relative humidity experienced by the sample.\cite{FALK:1968p982}

A comparison of the radial momentum distributions,  4$\pi p^2$n(p), of the protons in DNA with(Hydrated) and without the hydrated water molecules(Dry) is shown in Fig 1. The values of $\sigma$ and the significant parameters in the fits are shown in Table 1 for the hydrated and dry  samples.  The error bars shown in the graphs are calculated from the least squares fitting procedure, and include the correlations between the parameters. Terms up to n=7 were included in the initial fits. The parameters set to zero were found to have average values less than half their rms variance, and did not significantly raise the chi-square of the fit when set to zero. Their inclusion enhances the rms point by point error, but has a negligible affect on the values of n(p).\cite{Andreani05}  
\begin{figure}[h] 
   \centering
   \includegraphics[width=5in]{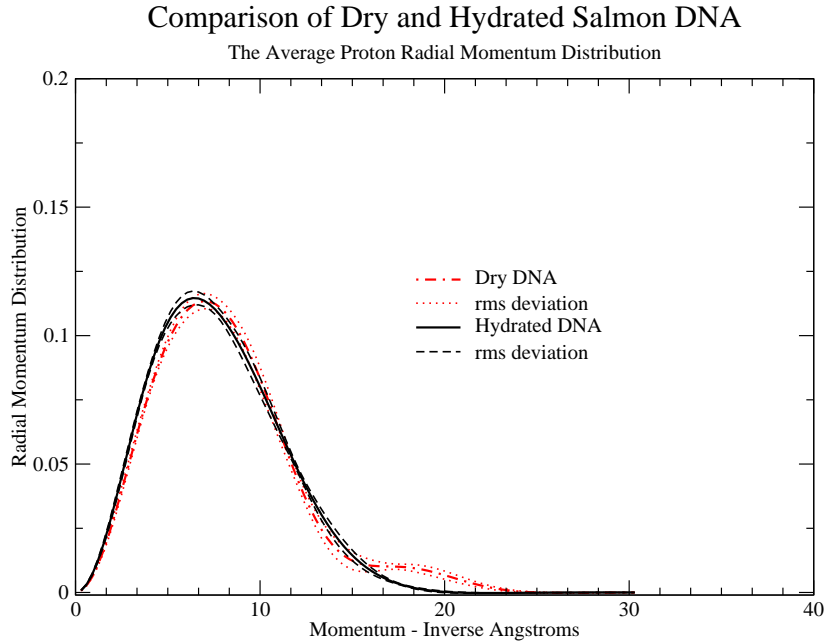} 
   \caption{Comparison of the average momentum distribution of the protons in hydrated Salmon DNA with that in dry Salmon DNA. The oscillation in the distribution of the dry sample indicates coherent delocalization of some of the protons in a double well. The position of the local minimum allows an estimation of the separation of the wells as .2$\AA$ }
   \label{fig:data}
\end{figure}
\begin{table}[h]
\centering
 
\caption{The fitting coefficients and their variances for
the momentum distribution measured in dry and hydrated DNA}
\begin{small}
\begin{tabular}{|c|c|c|c|c|}
\hline
\hline
\multicolumn{5}{c}{Parameters of measured n(p)}     \\ 
 \hline
\hline
 & \multicolumn{2}{c}{ Hydrated}\vline&\multicolumn{2}{c}{Dry}\vline \\
  \hline
$\sigma$ ($\AA$$^{-1}$) &\multicolumn{2}{c}{4.81$\pm$  .028 }\vline&\multicolumn{2}{c}{ 5.19 $\pm$ .067 }\vline\\
\hline
$n$ & $a_{n}$ & $\delta a_{n}$& $a_{n}$ & $\delta a_{n}$\\
\hline
2 & 0 & - &  .657 & .134\\
3 &-.259 &.075  & 0 & -\\
4 & 0 & - &  -.438 & .177\\
5 &0 & -&0& -\\
6 & .086 & .101 &  .354 & .178\\
\hline

\end{tabular}
\end{small}
\end{table}
We can see immediately that the dry sample has a larger second moment, as is confirmed by the values of $\sigma$ in Table 1.Moreover, n(p) is not monotonic for large p. The appearance of an oscillation in the momentum distribution is characteristic of  a population of protons that are coherently distributed in the ground state of a double well potential. The bimodal spatial wavefunction, when fourier transformed to give a momentum wave function, produces oscillations with a wavelegth in momentum space of $\pi$/d \cite{RS}, where d is the separation of the wells.In this case, we obtain a value of d of .2$\AA$ 

  Evidently, the hydration of DNA, at least for the first 6 molecules of water, is accompanied by a reduction of the zero point kinetic energy of the water-DNA system. That energy is available to produce a distortion of the A phase of the DNA towards the B phase. We wish to see how significant it is compared to the enthalpy change per water molecule upon removing the same number of water molecules from the DNA and placing them in bulk water. This has been measured by K. Virnik et al.\cite{virnik:2001p1026} and others \cite{Hoyer:1981p1023,cavanuaugh:2002p1018} using Differential Scanning Calorimetry.The value is 3.0$\pm .15$ KJ/mol for salmon DNA at this level of hydration. To do this, we compare the total kinetic energy of the dry DNA together with  6 water molecules in bulk water, with that of the hydrated DNA. 
  
  That is, if the number of protons in the dry sample is N$_{DNA}$ , and the number of protons in the water that was removed from the hydrated sample is N$_{H_2O}$, then in the hydrated sample the total kinetic energy is (N$_{H_2O}$+N$_{DNA})\hb\sigma^2_{Wet}$. The kinetic energy of the dry sample is N$_{DNA}\hb\sigma^2_{Dry}$ to which must be added the kinetic energy of the water molecules in bulk water, N$_{H_2O}\hb\sigma^2_{H_2O}$. The change in the kinetic energy/proton, averaged over all the protons,  from the dry state to the hydrated state is therefore.
\begin{equation}
\label{ke}
\Delta K.E.=\hb[\nd\sigma^2_{Dry} +\nw\sigma^2_{H_2O}-(\nd+\nw)\sigma^2_{Wet})]
\end{equation}
The change in the kinetic energy per proton is therefore
\begin{equation}
\label{ke2}
\Delta K.E/N_{Tot}.=\hb[\frac{\nd}{\nt}(\sigma^2_{Dry}-\sigma^2_{Wet}) +\frac{\nw}{\nt}(\sigma^2_{H_2O}-\sigma^2_{Wet}]
\end{equation}
where N$_{Tot}=\nd+\nw$. The sigma parameter for bulk water was measured at the same time as the other experiments descrived here. The value of  $\sigma$=4.76$ \pm .035 \AA$ is consistent with earlier measurements.\cite{Pantalei:2008p963}. With the ratio of protons in the dry DNA to that in the hydrated  DNA as calculated from the stoichiometry as .696, and the values for the $\sigma$ parameters and their uncertainty given in Table 1, we find the average change in the kinetic energy/water molecule,   to be, since there are two protons in a water molecule
\begin{equation}
\label{ke2}
2\Delta K.E/N_{Tot}.=3.02\pm .51 KJ/mol
\end{equation}
That is,  the change in kinetic energy of the protons in the combined water-Dna system, in going from the dry phase, with the water molecules in bulk water and the DNA  unhydrated, to the hydrated phase, completely accounts for the measured enthalpy change. If there is some water still in what we have been calling the dry phase, the change in kinetic energy/proton would be increased abve the value given above. This does not mean that the electrostatic interactions are insignificant compared to the kinetic energy changes, only that the reduction  in the electrostatic potential energy from the incorporation of the molecules of water in the DNA is nearly equal to the increase in the elastic potential energy of  the A phase as it deforms into the B phase.  The reduction in zero point energy then provides  the binding energy for the water molecules. 

The extent to which this change in zero point energy can be attributed to the protons in the hydrating water, rather than the hydrogen bonds in the DNA  is unknown. Comparable changes in the kinetic energy of protons in water molecules have been seen in water confined in carbon nanotubes\cite{nanoprl} and 
in xerogel\cite{Garb1}, so one might think most of the change  is due to changes in the environment of the protons in the water molecules. The shape of the momentum distribution for the dry phase, however,  suggests changes in the base pair  hydrogen bonds, as the potentials for the protons there are sensitive to the separation of the base pairs, and this is easily changed with a structural deformation.  It is also the case that the measured changes in the stretch and bending modes vibrational frequencies for the hydrated water,  which account for most of the zero point energy, are far too small( at most 1.5$\%$) to account for the changes in the kinetic energy if we assume the protons in the water remain in approximately harmonic potential wells.\cite{FALK:1964p972,TAO:1989p985}  In any case, our results show that a correct description of the energetics of hydration of DNA requires that the quantum delocalization of the protons in the system be accounted for. We expect  this will be generically true of other systems in which the hydrogen bond network of water is strongly distorted, due for instance, to the confinement of the water in cells.

\centerline{Acknowledgements}
This work was supported by the DOE, Office of Basic
Energy Sciences under Contract  No.DE-FG02- 03ER46078, and  partially supported within the CNR-CCLRC Agreement No. 01/9001 concerning collaboration in scientific research at the spallation neutron
source ISIS. The financial support of the Consiglio Nazionale delle Ricerche in this research is hereby acknowledged. G. Reiter wishes to thank Alan Mills and Phil Platzman and William Widger for useful converstions, and William Widger for guidance and assistance in preparing the samples. . 
\bibliography{bib}

\end{document}